# Pressure induced electride phase formation in calcium: A key to its strange high-pressure behavior

P. Modak*and Ashok K. Verma
*High Pressure & Synchrotron Radiation Physics Division, Bhabha Atomic Research Centre, Mumbai-400085, India*

Peter M. Oppeneer
*Department of Physics and Astronomy, Uppsala University, P. O. Box 516, SE-75120 Uppsala, Sweden*

Elemental calcium (Ca), a simple metal at ambient conditions, has attracted huge interest because of its unusual high-pressure behavior in structural, electrical, and melting properties whose origin remain unsolved. Here, using a theoretical framework appropriate for describing electride phase formation, i.e., the presence of anionic electrons, we establish electride formation in Ca at a pressure as low as 8 GPa. Our analysis shows that under pressure the valence electrons of Ca localize at octahedral holes and exhibit anionic character which is responsible for its strange pressure behavior. Our calculated enthalpy and electrical resistance indicate that Ca will directly transform from an FCC-electride phase to an SC-electride phase near 30 GPa thereby avoiding the intermediate BCC phase. These findings are not limited to Ca but might hold a key to the understanding of host-guest type structures which occur in other elemental solids though at much higher pressures.

Calcium exhibits strange high-pressure behavior in the 8-50 GPa pressure region [1-12]. The room-temperature electrical resistance of Ca shows a continuous increase from 8 to 18 GPa then it gradually decreases attaining a lowest value near 25 GPa and afterwards it again starts increasing, attaining the highest value near 42 GPa which is highly unusual for a good metal like Ca [2]. Subsequent x-ray diffraction (XRD) measurements advocated a face-centered cubic (FCC) to body-centered cubic (BCC) structural transition near 19.2 GPa and then a BCC to simple cubic (SC) transition near 33 GPa which is however contrary to the widely accepted high-pressure structural behavior of most materials that are expected to adopt progressively more and more close-packed structures under compression [3]. The occurrence of BCC structure near 19.2 GPa is also abnormal as BCC is a high-temperature ambient-pressure phase which has a good metallic nature at all pressures and hence this phase cannot explain the observed resistance behavior [2]. The experimental assignment of BCC phase, near 19.2 GPa, is based on the fact that a single equation-of-state could be adopted for both an unknown phase and ambient-pressure high-temperature BCC phase [3]. A recently claimed BCC observation was also not totally convincing [6] since the most prominent peak, indexed as (111), for BCC-Ca in the XRD pattern will have zero value for the structure factor. Further, all high-pressure experiments were carried out under non-hydrostatic conditions which may strongly influence the phase transition kinetics. A large transition pressure mismatch between the experimental and theoretical results was furthermore reported for the FCC to BCC transition [3-7, 10-12]. Moreover, the experimental observation of SC-Ca is controversial as quasi-harmonic phonon calculations using semi-local functionals gave imaginary phonon frequencies for SC-Ca at 0K [8, 9, 12].

Subsequent crystal structure searches found a tetragonal $I4_1/amd$ (BCT) structure which becomes energetically favorable over SC near 34 GPa [10]. Although several theoretical approaches have been applied to resolve the structural inconsistency between theory and experiments, a consistent phase diagram of Ca is still missing in the 0-50 GPa region [11-15].

Further, Ca was proposed to adopt a host-guest type structure above 134 GPa [16]. It is worth mentioning that startling complexities were observed in simple elemental solids under high-pressure. Some elements adopt a non-periodic incommensurate/host-guest structure; in some cases, the elements show spectacular changes in physical properties at high-pressures. Barium metal, for example, adopts a highly complex host-guest structure near 12 GPa and metals like Li, Na, K, Mg, and Cs turn into a semimetal or insulator when subjected to extreme high-pressures [17-22]. It is widely believed that under extreme pressures these form electrides, an unusual class of materials containing localized electrons at the non-nuclear sites, i.e., at interstitial sites which act like an anion [17-21, 23, 24]. There are several alkali and alkali-earth metals containing compounds which host electride phases at ambient conditions [25-29] but an elemental electride is unknown. Identification of an electride in Ca may allow experiments to explore the role of anionic electrons for complex phase formation in an elemental solid at low pressures.

In this Letter, using a hybrid exchange-correlation functional known to give correct descriptions for localized electrons, we explore the possibility of electride formation in Ca and establish the formation of such phase at 8 GPa. Since experimental detection of the interstitial anionic electrons is not straight forward under high-pressures, electron density-based descriptors like Bader analysis [30], electron-localization-function (ELF) [31], and presence of non-nuclear maxima and their charge content are usually used to detect the electride phase in solids. We adopt this methodology and obtain clear indicators of a Ca electride phase at 8 GPa. Our results demonstrate that a BCC phase would not exist in Ca at high-pressures at 0K when we correctly treat its localized anionic electrons. Also, the calculated resistance for electride FCC-Ca reproduces the measured high-pressure electrical resistance data satisfactorily, explaining Ca's strange resistivity behavior.

To establish pressure-induced electride formation, we calculated total energy, ELFs, Bader charges, and band decomposed charge density using the Vienna Ab Initio Simulation Package (VASP) for different phases at different pressures [32-34]. The projector augmented wave potential with configuration $3s^2 3p^6 4s^2$ is used in conjunction with 346.6 eV effective cut-off for plane wave basis set [35]. For exchange-correlation, the generalized gradient approximation (PBE-GGA) and Heyd-Scuseria-Ernzerhof (HSE06) hybrid functional are used [36, 37]. It is worth mentioning, in connection to the localization of valence electrons, that large over-delocalization errors are expected for semi-local density functionals such as GGA whereas hybrid functional are expected to give a proper description of such systems. A 36×36×36 k-point grid is used for Brillouin zone sampling for the cubic structures and equivalent Δk (~ 0.03 Å$^{-1}$) is used for non-cubic structure. Electronic charge density critical points are calculated using Critic2 program [38]. A 4×4×4 supercell is used for phonon calculations [39].

In Fig.1, we present our pressure-enthalpy data of the FCC, BCC, SC, and BCT phases of Ca. The obtained zero-pressure properties with both types of functionals compare fairly well with previous theoretical and experimental results (see Table 1S [40]). However, it is worth mentioning that the HSE06 gives a much better estimation of these quantities. Analogous to earlier studies, GGA functional gives an FCC to BCC transition near 8.2 GPa and BCC to BCT transition near 27.6 GPa and there is no sign of a stable SC phase up to 50 GPa [10, 11, 16], contrary to what has been observed in the experiments [3-6]. This excellent matching with earlier theoretical works [10, 14, 16] establishes the reliability of our calculations and chosen parameters like energy cutoff, k-point grid etc. Remarkably, HSE06 predicts an FCC to SC transition near 30 GPa whereas BCC and BCT structures remain unfavorable up to 50 GPa (see Fig.1b). This contrasting phase stability behavior compared to GGA may results from the inability of GGA functional to treat localized charge density which was predicted to exist in Ca but at high pressure [16]. It is important to note that an earlier study with a hybrid functional has not shown the FCC stability region but gave a stable BCC phase in the 20-28 GPa pressure region, a BCT phase in the 28-33 GPa pressure region and a SC phase above 33 GPa [15]. Furthermore, our computed enthalpy curves reveal that the BCC phase might stabilize in the negative pressure region, consistent with the fact that BCC has been observed at ambient-pressure, high-temperature conditions [41].

To examine electride phase formation in Ca we analyzed ELF, a strong indicator for an electride phase. It takes values between 0 and 1.0 where value 1.0 corresponds to the perfect localization limit and value 0.5 represents the homogeneous electron gas limit which is best describe by semi-local functionals such as LDA and GGA [42, 36]. Thus, systems containing high ELF values are prone to an over-delocalization error when treated by semi-local functionals.

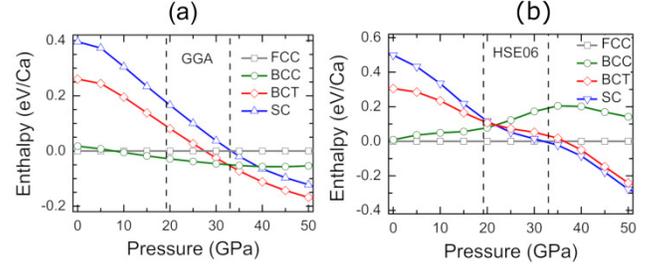

Fig. 1. Calculated high-pressure solid-solid structural phase transitions in Ca. (a) & (b) Pressure variation of enthalpy for BCC, BCT, and SC phases relative to the ambient FCC phase. Vertical dash line represents experimental phase boundaries (FCC to BCC at 19.2 GPa and BCC to SC at 33 GPa) at room temperature.

Figs. 2(a)-(b) present ELF data for FCC and SC structures. It is evident that for these structures the highest ELF occurs at non-nuclear interstitial sites (octahedral holes). For the GGA functional, we find that the ELF value increases from 0.70 to 0.85 for FCC structure in 0-21 GPa pressure interval and on further compression the ELF value gradually decreases (Fig. 2(c)). With HSE06 functional, the highest ELF value varies from 0.85 to 0.90 in the 8-22 GPa pressure region. For the SC phase, the highest ELF value varies from 0.85 to 0.92 in case of GGA and 0.93 to 0.95 in case of HSE06 over the 20-45 GPa pressure interval (Fig. 2(d)). For the BCC structure, the ELF has comparatively lower values and, importantly, the highest value 0.70 occurs in the vicinity of atomic sites and it remains practically independent of the pressure for both functional (shown in Fig. S1(a) [40]). Notably, the BCC structure exhibits no interstitial charge localization at any pressure. The BCT structure too exhibits interstitial electron localization (shown in Fig. S1(b) [40]) and the highest ELF value varying from 0.80 to 0.82 for HSE06 in the pressure region of 20-35 GPa. The occurrence of interstitial charge localization in FCC, SC and BCT structures indicates the likelihood of electride formation under favorable conditions. The ELF analysis explicitly establishes further that a large electron over-delocalization error is unavoidable for FCC and SC structures when using semi-local functionals since these structures possess a very high

value of ELF in their octahedral holes. Thus, semi-local functionals used in earlier studies have underestimated the FCC stability region. It is also highly possible that the over-delocalization error of semi-local functionals has contributed to the predicted dynamical instability of the SC structure whose existence was however convincingly shown by several experiments [3-6].

Next, we analyze the critical points (CPs) of the crystal electron charge density characterized by zero gradient of the electron density [43]. There are four types of CPs, namely: local minimum or cage CP (ccp), local maximum or nuclear CP (ncp) and two kinds of saddle points, bond CP (bcp) and ring CP (rcp).

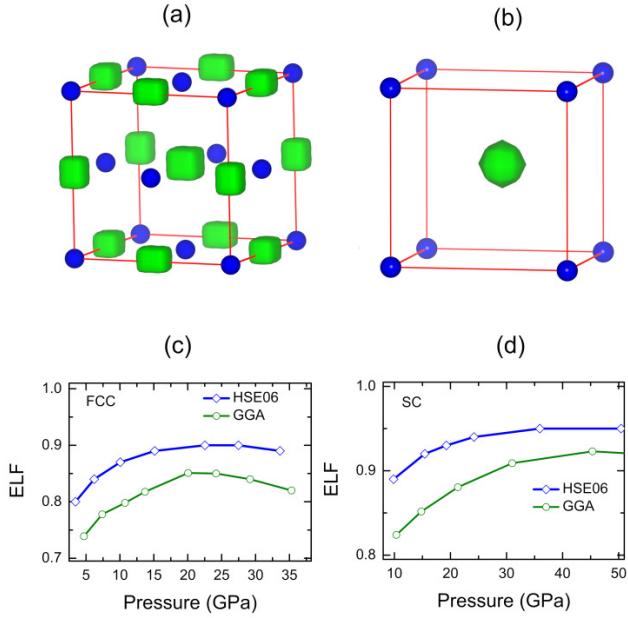

Fig.2. Computed ELFs of FCC-and SC-Ca for HSE06 functional. (a) FCC-Ca ELF surface for an iso-value=0.80, near 10 GPa, (b) SC-Ca ELF surface for an iso-value=0.88, near 32 GPa. Here, small blue balls represent Ca atoms and ELF surfaces are shown in green color. (c) & (d) Pressure dependence of the highest ELF values of FCC-and SC-Ca, respectively.

A charge density minimum (maximum) is associated with positive (negative) curvatures of the field lines in all 3-directions, bcp has two negative curvatures and rcp has one negative curvature out of three. The occurrence of a charge density local maximum in the interstitial region, i.e., a non-nuclear maximum (NNM) is another strong indicator of an electride phase formation provided it has significant charge content (>10% of the valence charge) [44]. We estimated the NNM charge content using the Yu-Trinkle algorithm, as implemented in the Critic2 program [45, 38]. We found that the FCC phase has NNMs at both tetrahedral holes and region surrounding octahedral holes below 8 GPa (shown in Fig. S2 [41]). However, these NNMs have relatively small charge content, $0.004e^-$ at tetrahedral holes and $0.02e^-$ around octahedral holes. We observed that close to 8 GPa the NNM at tetrahedral holes vanish and NNMs around the octahedral hole merge and move right at the octahedral position (shown in Fig. 3a) which was also reported in anearlier study [46]. Charge content of this NNM is $0.25e^-$ for GGA and $0.60e^-$ for HSE06. This confirms the interstitial anion formation which is also known as interstitial quasi atom (ISQ). Thus, our analysis establishes electride phase formation in the FCC structure. Furthermore, it is evident from Fig. 3a that the bond CPs are located between Ca and ISQ sites indicating dominant interactions between Ca and ISQ. It is pertinent to mention that an earlier study named this transition an isostructural transition [46]. However, it is apparent from our analysis that the electride phase formation represents a phase transition from FCC to a NaCl (B1) type pseudo-binary phase, i.e., an ionic solid-like behavior above 8 GPa. On further compression, the charge content of this ISQ increases reaching a value of $0.46e^-$ for GGA and $0.71e^-$ for HSE06 near 20 GPa and afterward its charge content starts decreasing (shown in Fig. 3c). This is consistent with the observed electrical resistance behavior [2].

From a similar analysis, we find that the SC structure also has NNM at octahedral holes and their charge content near 30 GPa is $0.52e^-$ for GGA and $0.81e^-$ for HSE06 (shown inFig. 3b). The NNM charge content gradually decreases with pressure (shown in Fig. 3d). This establishes that the SC phase also forms an electride phase, analogue to a pseudo-binary CsCl (B2) type solid. Remarkably, the BCC structure does not have NNMs whereas the BCT structure has NNMs but their charge content is exceptionally small. Thus, we anticipate that Ca in BCC and BCT structures will exhibit good metal behavior. Itis further worthwhile to mention the striking resemblance of the high-pressure melting behavior of Ca and NaCl solid [47, 48] where both materials show a nearly flat melting curve in the 10-30 GPa pressure interval and then exhibit a sudden rise in melting temperature near 30 GPa, a pressure value which is close to the B1-B2 transition of the NaCl solid.

Existence of anionic electrons in a solid usually results in a high-lying partially occupied state in its electronic bandstructure. For FCC and SC phases, this state hybridizes with the atomic $s$-state and lies close to the Fermi level ($E_F$). Moreover, this state pushes up the $3d$-states, away from the Fermi level as evident from the electronic bandstructure with HSE06, shown in Figs.4(a) & (b) and thus reducing the possibility of $3d$ occupancy. Therefore, our findings strongly indicate that the long-held pressure induced $s$ to $d$ electron transfer in Ca is not correct. Additionally, to support the presence of electride state we have calculated band-decomposed charge density which evidently shows that the partially occupied band, as indicated by red color in Figs. 4(a) & (b), is associated with the anionic electrons that are shown in Figs. 4(c) & (d). To show the electride state in a further way we have calculated TB-LMTO [49] bandstructure by putting an empty sphere (E) which acts as a replica of ISQ, at the octahedral hole of both phases. The fat band plots, shown in Fig. S3 [40], are associated with the empty sphere which, too, proves the presence of an electride state.

To gain insight in chemical bonding in the electride phases of Ca, we have computed the crystal orbital Hamiltonian population (COHP) [50] and its integrated value (ICOHP) under pressure for FCC and SC phases using TB-LMTO code [49]. Negative COHP values indicate a bonding interaction whereas positive values indicate an anti-bonding interaction and the ICOHP up to the Fermi energy gives an estimate of the bond strength [50].

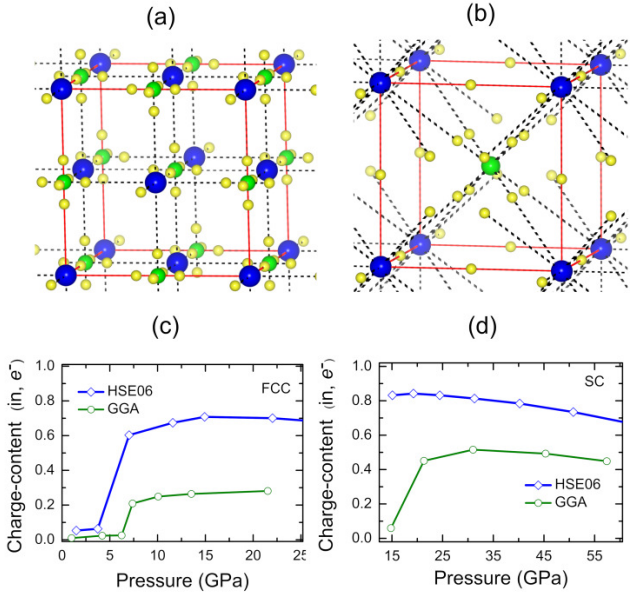

Fig. 3. NNM and bcp of Ca under pressure calculated using HSE06 functional. (a) NNM at octahedral holes (green balls), and bcp (yellow balls) in FCC-Ca near 15 GPa. (b) NNM (green balls) and bcp (yellow balls) in SC-Ca near 34 GPa. Here, blue balls represent Ca atoms. (c) & (d) Pressure dependence of NNM charge content at octahedral holes in FCC and SC phases.

For both phases, we find that the Ca-Ca pair has much smaller COHP and ICOHP values in comparison to that of Ca-E pair, indicating an effective chemical bonding between Ca and ISQ (shown in Figs. S4a & b [40]). This agrees with our charge density critical point analysis where we found bond paths between Ca and ISQ sites. This analysis further supports that the chemical bonding in Ca metals above 8 GPa is through interactions between $Ca^+$ at atomic sites and anionic electrons located at octahedral holes, *i.e.*, effectively an ionic interaction analogous to $Na^+$ - $Cl^-$ in the FCC and $Cs^+$ - $Cl^-$ in the SC structure. For both phases the chemical bond strength shows an increasing trend with increase of pressure (see Figs. S4c & d [40]).

To examine dynamical stability of the electride SC-Ca structure we carried out phonon calculations with HSE06. We found dynamically stable SC-Ca at 0K above 30 GPa (shown in Fig.S5 [40]) contrary to earlier calculations with semi-local functional which reported imaginary phonon frequencies for SC-Ca at high-pressure conditions at 0K, thereby questioning previously the existence of SC-Ca [8-9, 12].

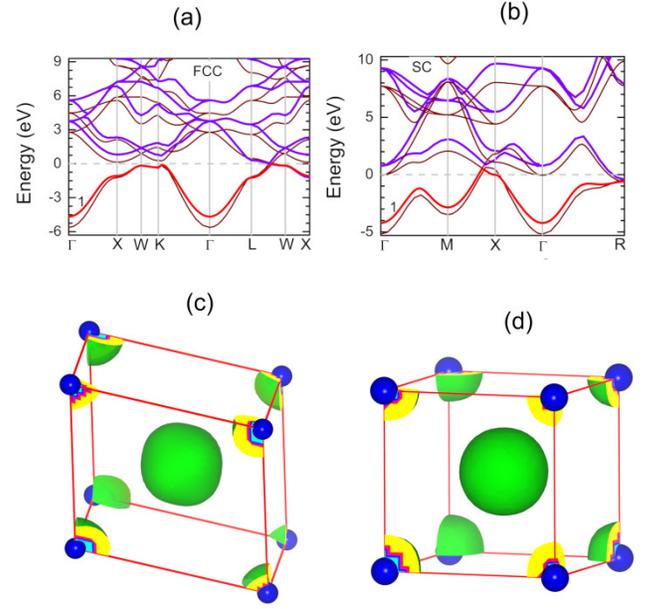

Fig. 4. Electronic bandstructure and electride charge density surfaces. (a) FCC-Ca bandstructure near 14 GPa, (b) SC-Ca bandstructure near 40 GPa. Here, thin wine curves represent GGA bands whereas thick violet and red curves represent HSE06 bands. (c) FCC-Ca HSE06 band decomposed charge density surface for an iso-value = $0.07e^-$. (d) As in (c), but for SC-Ca for an iso-value = $0.15e^-$.

Next, to understand the observed electrical resistance behavior, we used Bloch resistivity formula, considering electrons having parabolic dispersions and scattered by Debye phonons [51]. It is to be noted that the Bloch formula accounts only for normal processes of electron-phonon scattering. Except at very low temperature an umklapp process may become important. The probability of an umklapp process depends both on size of the Fermi sheets and the distance between two sheets in neighboring zones. From our calculated bandstructure of FCC-Ca near 14 GPa (Fig. 4a) it is evident that partially occupied bands cross the Fermi level along L-W direction of BZ where L-point is the centre and W-point is the corners of the hexagonal face of this zone. Thus, a ring like narrow Fermi sheet around L-point on this hexagonal face is expected for FCC-Ca and also, Fermi sheets at two neighboring zones will be far apart as they occur at zone boundary. Therefore the probability of umklapp scattering is expected to be small because a phonon of higher energy is needed to drive the electron over a longer distance to land on another Fermi surface in a neighboring BZ.

Assuming isotropic compression one can convert the resistivity into resistance and then the normalized resistance can be written as

$$\frac{R_P}{R_0} = \left(\frac{V_P}{V_0}\right)^{3/2} \left(\frac{\theta_D^0}{\theta_D^P}\right)^6 \frac{I_5(x_P)}{I_5(x_0)} \frac{N_0(E_F)}{N_P(E_F)}, \quad (1)$$

where $x = \theta_D/T$ and $I_5(x) = \int_0^x \frac{z^5 e^z}{(e^z-1)^2} dz$. Here '0' refers to quantities at zero pressure and 'P' refers to corresponding quantity at pressure $P$ GPa. $N(E_F)$ is the DOS at $E_F$ per unit cell having volume $V$. Since Ca bands close to $E_F$ have considerable dispersions, we can safely use this formula. We calculated the Debye temperature $\theta_D$ at different pressures and temperatures by computing the phonon DOS and then taking second order moment of the phonon frequencies and are presented in Table S2 [40]. Fig.5 shows the pressure variation of normalized resistance for GGA and HSE06 functionals along with the experimental data. To check the effect of electronic temperature on resistance we calculated the electronic DOS using Fermi-Dirac smearing with widths of 100 K and 300 K and also Methfessel-Paxton smearing with width 0.1 eV for GGA (shown in Fig. S6a [40]). It is evident that GGA can explain the first peak in the observed pressure variation of resistance but fails above 20 GPa. The right shift of the first peak with temperature is also reproduced by GGA. As the pressure variation of resistance is nearly identical for Fermi-Dirac smearing corresponding to 300 K and Methfessel-Paxton smearing with width 0.1 eV, we used the latter one for computing electronic DOS with the HSE06 (see Fig. S6b [40]). The computed pressure variation of resistance for HSE06 agrees well with that of the observed behavior at 300 K. It is evident from Fig.5 that the normalize resistance of BCC-Ca does not follow the observed behavior indicating an improper assignment of this structure in the pressure range of 0-50 GPa.

In summary, we have shown that electride phase formation is responsible for the strange high-pressure behavior of Ca. Under compression Ca forms an FCC NaCl-type electride near 8 GPa, which transforms to another SC CsCl-type electride phase near 30 GPa thereby completely skipping a BCC phase. Our analysis has shown that semi-local functional yield a large over-delocalization error, which strongly affects the predicted high-pressure behavior, particularly the energetics of high-pressure phases. Hence, functionals that capture electron localization, such as hybrid functionals are essential to study the high-pressure behavior of Ca. Furthermore, the computed electrical resistance with the HSE06 functional for FCC-Ca shows satisfactory agreement with measured resistance data up to 30 GPa. Our findings are expected to be relevant for other alkali and alkali-earth metals that exhibit highly complex crystal structures at high-pressure conditions where the formation of such crystal lattices demands the existence of a two-component solid.

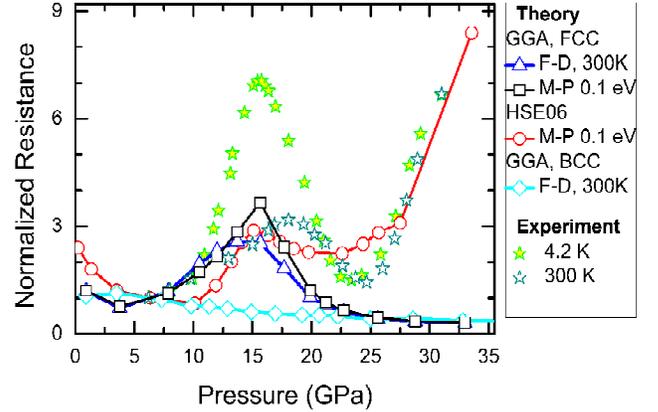

Fig. 5. Pressure variation of electrical resistance of Ca up to 35 GPa. Experimental data are from Ref. 2. Here, F-D stands for Fermi-Dirac smearing and M-P stands for Methfessel-Paxton smearing methods.

**Acknowledgment**

P.M. and A.K.V. acknowledge Computer Division, BARC for providing the supercomputing facility. P.M.O. acknowledges support from the Swedish Research Council (VR).

*Corresponding author.

**References**:
[1] H. Fujihisa et al., Phys. Rev. Lett. **110**, 235501 (2013).
[2] K. J. Dunn and F. P. Bundy, Phys. Rev. B **24**, 1643 (1981).
[3] H. Olijnyk and W. B. Holzapfel, Phys. Lett. **100A**, 191 (1984).
[4] T. Yabuuchi, Y. Nakamoto, K. Shimizu, and T. Kikegawa, J Phys. Soc. Jpn. **4**, 2391 (2005).
[5] Q. F. Gu, G. Krauss, Y. Grin, and W. Steurer, Phys. Rev.B**79**, 134121 (2009).
[6] S. Anzellini et al., Phys. Rev. Mater. **2**, 083608 (2018).
[7] S. Arapan, Ho-kwang Mao, and R. Ahuja, Proc. Natl. Acad. Sci. USA **105**, 20627 (2008).
[8] G. Gao et al., Solid State Commun. **146**, 181 (2008).
[9] I. Errea, M. Martinez-Canales, A. R. Oganov, and A. Bergara, High Press. Res. **28**, 443 (2008).
[10] Y. Yao, D. D. Klug, J. Sun, and R. Martoñák, Phys. Rev. Lett. **103**, 055503 (2009).
[11] A. M. Teweldeberhan and S. A. Bonev, Phys. Rev. B **78**, 140101 (2008).
[12] T. Ishikawa et al., Phys. Rev. B **77**, 020101 (2008).

[13] J. S. Tse, S. Desgreniers, Y. Ohishi, and T. Matsuoka, Sci. Rep. **2**, 372 (2012).
[14] M. Di Gennaro, S. Kumar Saha, and M. J. Verstraete, Phys. Rev. Lett. **111**, 025503 (2013).
[15] H. Liu, W. Cui, and Y. Ma, J. Chem. Phys. **137**, 184502 (2012).
[16] A. R. Oganov, Y. Ma, Y. Xu, I. Errea, A. Bergara, and A. O. Lyakhov, Proc. Natl. Acad. Sci. USA **107**, 7646 (2010).
[17] R. J. Nelmes, D. R. Allan, M. I. McMahon, and S. A. Belmonte, Phys. Rev. Lett. **83**, 4081 (1999).
[18] T. Matsuoka and K. Shimizu, Nature **458**, 186 (2009).
[19] Y. Ma et al., Nature **458**, 182 (2009).
[20] M. Marqués et al., Phys. Rev. Lett. **103**, 115501 (2009).
[21] M.–S. Miao and R. Hoffmann, J. Am. Chem. Soc. **137**, 3631 (2015).
[22] K. Takemura et al., Phys. Rev. B **61**, 14399 (2000).
[23] J. L. Dye, Science **247**, 663 (1990).
[24] H. G. von Schnering and R. Nesper, Angew. Chem., Int. Ed. Engl. **26**, 1059 (1987).
[25] S. Matsuishi et al., Science **301**, 626 (2003).
[26] M. Hirayama et al., Phys. Rev. X **8**, 031067 (2018).
[27] D. Gregory, A. Bowman, C. Baker, and D. Weston, J. Mater. Chem. **10**, 1635 (2000).
[28] R. H. Huang et al., Nature **331**, 599 (1988).
[29] D. J. Singh, H. Krakauer, C. Haas, and W. E. Pickett, Nature **365**, 39 (1993).
[30] R. F. W. Bader, Chem. Rev. **91**, 893 (1991).
[31] F. Hao, R. Armiento, and A. E. Mattsson, J. Chem. Phys. **140**, 18A536 (2014).
[32] G. Kresse and J. Furthmüller, Phys. Rev. B **54**, 11169 (1996).
[33] G. Kresse and J. Hafner, Phys. Rev. B **47**, 558 (1993).
[34] G. Kresse and J. Furthmüller, Comput. Mater. Sci. **6**, 15 (1996)
[35] P. E. Blöchl, Phys. Rev. B **50**, 17953 (1994).
[36] J. Perdew, K. Burke, and M. Ernzerhof, Phys. Rev. Lett. **77**, 3865 (1996).
[37] A. V. Krukau, O. A. Vydrov, A. F. Izmaylov, and G. E. Scuseria, J. Chem. Phys. **125**, 224106 (2006).
[38] A. Otero-de-la-Roza, E. R. Johnson, and V. Luãna, Critic2: Comput. Phys. Commun. **185**, 1007 (2014).
[39] D. Alfè, Comput. Phys. Commun. **180**, 2622 (2009). *URL*: http://chianti.geol.ucl.ac.uk/~dario
[40] See Supplemental Materials for a table containing ambient pressure properties, ELF iso-surfaces for BCC and BCT structures, NNM for FCC-Ca below 8 GPa, LMTO bandstructure character plots for FCC-Ca, COHP functions and pressure variation of ICOHP for both FCC and SC Ca and pressure variation of DOS at $E_F$ for FCC, SC and BCC structures.
[41] J. F. Cannon, J. Phys. Chem. Ref. Data **3**, 791 (1974).
[42] J. P. Perdew and Y. Wang, Phys. Rev. B. **45**, 13244 (1992).
[43] V. G. Tsirelson, Can. J. Chem. **74**, 1171 (1996).
[44] S. G. Dale and E. R. Johnson, J. Phys. Chem. A **122**, 9371 (2018).
[45] M. Yu and D. R. Trinkle, J. Chem. Phys. **134**, 064111 (2011).
[46] T. E. Jones, M. E. Eberhart, and D. P. Clougherty, Phys. Rev. Lett. **105**, 265702 (2010).
[47] D. Errandonea, R. Boehler, and M. Ross, Phys. Rev. B **65**, 012108 (2001).
[48] R. Boehler, M. Ross, and D. B. Boercker, Phys. Rev. Lett. **78**, 4589 (1997).
[49] O. K. Andersen, Stuttgart Tight-binding LMTO Program version 4.7, Max Planck Institutfür Festkörperforschung.
[50] R. Dronskowski and P. E. Blöchl, J. Phys. Chem. **97**, 8617 (1993).
[51] Alka B. Garg et al., J. Phys.: Condens. Matter **14**, 8795 (2002).